\documentclass[11pt,a4paper]{article}

\usepackage{amsmath,amssymb,mathtools,amsthm}
\usepackage{hyperref}
\usepackage{fullpage}
\usepackage{graphicx}
\usepackage{subfigure}
\usepackage{xspace}
\usepackage{authblk}

\newcommand{\pkgname}[1]{\textsc{#1}}

\newcommand{\GG}{\mathcal G}

\newcommand{\ER}{Erd\H{o}s-R\'enyi\xspace}
\newcommand{\WS}{Watts-Strogatz\xspace}
\newcommand{\BA}{Barab\'asi-Albert\xspace}

\newcommand{\PP}{\mathbb P}
\newcommand{\graph}{G}
\newcommand{\subgraph}{H}
\newcommand{\vertices}{V}
\newcommand{\edges}{E}

\newcommand{\marked}{\operatorname{S}}
\newcommand{\qss}{\mathcal{QSS}}
\newcommand{\qs}{\mathcal{QSS}}
\newcommand{\et}{\mathcal{T}}
\newcommand{\eff}{\operatorname{eff}}

\newcommand{\alg}{\operatorname{\textbf{Alg}}}

\newtheorem*{twr}{Theorem}

\theoremstyle{definition}
\newtheorem*{definition}{Definition}

\begin{document}

\title{Impact of the malicious input data modification on the efficiency of 
quantum spatial search}
\author[1,2]{Adam Glos\thanks{aglos@iitis.pl}}
\author[1]{Jaros\l{}aw Adam Miszczak}
\affil[1]{Institute of Theoretical and Applied Informatics, Polish Academy of
	Sciences, Ba{\l}tycka 5, 44-100 Gliwice, Poland}
\affil[2]{Institute of Informatics, Silesian University of Technology,
	Akademicka 16, 44-100 Gliwice, Poland}

\date{}


\maketitle
	\begin{abstract}
		In this paper we demonstrate that the efficiency of quantum spatial search can
be significantly altered by malicious manipulation of the input data in the
client-server model. We achieve this by exploiting exceptional configuration
effect on Szegedy spatial search and proposing a framework suitable for
analysing efficiency of attacks on quantum search algorithms. We provide the
analysis of proposed attacks for different models of random graphs. The
obtained results demonstrate that quantum algorithms in general are not secure
against input data alteration.
	\end{abstract}
	
\section{Introduction}\label{sec:introduction} 

\paragraph{Motivation} 
	
While the intensive research effort invested in the area of quantum computing is
fully justified by groundbreaking theoretical
developments~\cite{childs2004spatial,harrow2017quantum, bernstein2017post}, year
by year scientists have discovered new limitations of quantum computing
devices~\cite{van2013blueprint}. Quantum algorithms have been proved to be
susceptible to noise, which may falsify the results of the computation. This
fact motivated the development of the theory of quantum error-correcting codes.
Unitary operation decomposition provides numerous problems including
applications to hardware with fixed topology \cite{saeedi2011synthesis}. These
aspects started to play critical role after first commercial quantum computing
systems became available. Furthermore, hardware attacks, based on the security
holes of conventional electronics, have been discovered for quantum
cryptographic protocols~\cite{lydersen2010hacking}. The issues mentioned above
demonstrated that the theoretical security confirmed by the laws of physics in
the ideal environment could be deceptive in the real-world applications. For the
users it is important to be aware that quantum algorithms are inefficient for
some types of input data. One of such examples is quantum spatial search, which
is inefficient on 2D grid~\cite{childs2004spatial}. This demonstrates that the
quantum algorithms are not only unsuitable for some types of problems, but, even
for generally optimal quantum algorithms~\cite{chakraborty2016spatial}, it is
possible to construct input data rendering them slow.
	
These examples show that, while quantum algorithms may be a milestone
in computational theory, their practical implementation and security still need 
to be carefully
checked. From the computational complexity point of view, it remains unknown whether 
the existing quantum algorithms can provide the speed-up promised by the
theoretical results, especially when applied to large datasets. 
On the other hand, the ability to describe the types of data which render quantum algorithms
ineffective can be used to develop post-quantum protocols.

Furthermore, as the remote utilization of quantum computational resources plays 
vital role for the development of quantum technologies, one should consider 
various aspects of the functioning of quantum devices accessed remotely. 
Currently, quantum transmission protocols used for Universal Blind Quantum 
Computation~\cite{broadbent2009universal} can be used for transferring quantum 
data unknown to the server. Such protocols provide also correctness of the data and 
authentication of the transmission. However, they cannot be used for 
preventing the situation when maliciously prepared data are provided and result 
in the undesired behavior of quantum machine. This might lead, for example, 
resulting in denial of service effect. The analysis of the impact that the 
malformed data can have on the quantum computing infrastructure is still 
missing and this aspect of using quantum computers in the cloud deserves more 
attention.

\paragraph{Contribution}


In this paper study how the input data can influence the efficiency of quantum 
algorithms based on quantum spatial search. We demonstrate that the efficiency 
of quantum algorithms can be significantly diminished by manipulation of the 
input data. For the purpose of this paper we consider the following scenario. 
We have two parties, Alice and Bob. Alice prepares input data and sends them to 
Bob, who has for his disposal quantum computational resources. Alice requests 
from 
Bob some kind of processing eg. sorting a list or finding  its maximum. This 
situation naturally occurs in cloud computing scenario. If the input data are 
altered by a third party, Eve, then the computational process executed by Bob 
may require more resources then initially assumed. In this situation Bob might 
not be able to handle queries from other users. However, the client-server 
scenario we consider here can also include the possibility that Alice is 
responsible for sending data which are already malformed. This might result 
either from the purposeful acting of Alice or from the fact that the data 
contain instances of problems which are hard for quantum computers to handle.

As an example one can consider a sorting via classical algorithm QuickSort. The
algorithm achieves average-case performance $O(n\log (n))$, for number of
elements $n$. However, for simple fixed pivot choice, one can provide malicious
input data, which increase computation time up to $O(n^2)$. If no security is
considered, choosing such data may considerably increase the computational
resources required for running the algorithm. If the computer processes such
data, this may result in denial-of-service. Fortunately, there are known
solutions to this problem, for example, choosing pivot randomly.
	
Similar problems arise in the quantum spatial search, based on different models
of quantum walks. For the continuous-time quantum walk, the algorithm works very
well and securely on many graphs
\cite{chakraborty2016spatial,chakraborty2017optimal,glos2017vertices}, achieving
efficiency $O(\sqrt{n})$, for the graph order $n$. However, there are known
examples, such as a two-dimensional grid, for which the algorithm reaches linear
complexity only~\cite{childs2004spatial}. While the grid graph can be searched
much faster for the discrete coined quantum walk, it is known that specific
subgraphs of the searched graphs can form so-called \emph{exceptional
	configurations}~\cite{nahimovs2017adjacent} which strongly reduce the
efficiency.

In this paper we consider the formation of exceptional configurations as a
method of attacking quantum spatial search. We propose a framework enabling the
analysis of the attack efficiency based on the expected runtime of the
algorithm. We utilize this framework for selected families of random graphs. We
take into account different resources available for the attacker. We also
discuss the connection between our results and the security protocols used for
Universal Blind Quantum Computation.

\section{Preliminaries}\label{sec:preliminaries}

Let $\graph=(\vertices,\edges)$ be an undirected graph and let $\marked\subset
V$ be a set of searched vertices. A quantum spatial search algorithm is a
quantum walk, which after $t$ steps finds any marked $v\in \marked$ with
probability $p$. While the classical spatial search is known to have time complexity
$\Omega(n)$, it is possible to achieve $\Theta(\sqrt{n})$ for quantum
algorithms~\cite{childs2004spatial,ambainis2005coins,chakraborty2016spatial,glos2017vertices}. The complexity may depend on a chosen graph~\cite{childs2004spatial} or a chosen set $\marked$ \cite{nahimovs2015exceptional}. Throughout this paper we assume that the complexity of the quantum spatial search grows as $\Theta(n^\alpha)$ for some $\alpha>0$ which is commonly observed for many types of graphs.

For \emph{discrete coined quantum walk} (DCQW), the search problem with
multiple marked vertices can be a hard
task for some combinations of marked vertices, known as exceptional configurations~\cite{nahimovs2015exceptional,nahimovs2017adjacent,nahimovs2017probability}. The existence of an exceptional configuration is demonstrated in two steps. First, the existence of a special stationary state needs to be shown. Second, a bound on the probability needs to be determined, based on the stationary state. Recently, the class of connected subgraphs having the stationary state has been described, which solves the first step.
\begin{twr}[\cite{prusis2016stationary}]
	Let $\graph$ be an arbitrary graph. Let $H$ be its induced connected subgraph. $H$ contains a stationary state, if
	\begin{itemize}
		\item it is not bipartite, or
		\item it is $(\vertices_\subgraph^1,\vertices_\subgraph^2)$-bipartite satisfying 
			\begin{equation}
		\sum_{v\in \vertices_\subgraph^1} \deg_\graph(v)=\sum_{v\in \vertices_\subgraph^2} \deg_\graph(v),
		\end{equation}
		where $\deg_\graph$ is the degree in graph $\graph$. 
	\end{itemize}
\end{twr}
The exceptional configuration of order 2 (2EC) is a path of length 2, such that
the degrees of vertices in the original graph $G$ are equal. The exceptional
configuration of order 3 (3EC) is a triangle graph, or a path of length 3, such
that the degree of the middle vertex equals the sum of the degrees of the end
vertices in the original graph $G$. Wong has shown the equivalence between
coined and \emph{Szegedy model} (SzQW) \cite{szegedy2004quantum,wong2016direct},
which is a general quantum walk definable on an arbitrary directed graph. We are
going to focus on the latter, as it is more suitable for numerical analysis.

\section{Description of the attack}
\label{sec:qss-notation}

Let us assume that Alice sends to Bob a description of the oracle, for example
in the form of the quantum circuit. Bob executes Szegedy quantum search based on
the obtained oracle. As a result Bob sends to Alice a marked vertex, found by
the algorithm. In this situation Bob expects to take advantage of the
capabilities of quantum computer and obtain the result within time
$\Theta(t^\alpha)$.

Let us now assume that Eve has her own quantum resources and intercepts the
description of the oracle send by Alice. She executes the algorithm and finds
the vertex in time $\Theta(t^\alpha)$. Then she creates a description of new
oracle, in which additional vertices is marked, and sends it to Bob. If marked
vertices form an exceptional configuration, then Bob will need $\Theta(t^\beta)$
time to execute the algorithm, with $\beta>\alpha$. One should note that
modification of input data send by Eve cannot be corrected by using the error
correction code, as the modification are made on a logical level.

Bob may detect that the oracle was altered if the algorithm consumes more
resources that it was initially assumed. Additionally, we assume that Bob knows
that the oracle has been modified by marking the additional vertices. This is
the worst possible scenario from Eve's perspective. In this situation he can
modify the hyperparameters of his algorithm to limit the adverse effect. The
simplest strategy is to change the measurement time. By this he may reduce the
execution time to $\Theta (t^{\overline{\beta}})$, with
$\overline{\beta}<\beta$.

\subsection{Framework for quantifying the efficiency of attacks}

For the purpose of quantifying the efficiency of attacking methods, we introduce the
formal description of the family of quantum spatial search algorithms.

\begin{definition}
	\emph{Quantum Spatial Search} $\qss $ is a tuple 
	$(\alg,t;\graph,\marked,\theta)$, where $\alg$ is a quantum algorithm 
	searching for any vertex $v\in \marked$ in time $t$, running on graph $G$,
	and parametrized by the set of parameters $\theta$.
\end{definition}
By $ p(\qs) $ we denote the success probability of $\qs$. Note that we do not
define $\theta$ precisely, as it depends on the chosen algorithm $\alg$. For
example, for coined quantum walk $\alg=\mathrm{DCQW}$, the parametrization
consists of the set of coin operators. For $\alg=\mathrm{SzQW}$, the parametrization
is the chosen stochastic operation $P$.

In order to compare different quantum spatial search algorithms we propose the following measure of efficiency.
\begin{definition}
\emph{Expected runtime} $\et_{\qss}$ of quantum spatial search
$\qss=(\alg,t;\graph,\marked,\theta)$ is defined~as
	\begin{equation}
	\et_{\qss} \coloneqq \frac{t}{p(\qss)}.
	\end{equation}
\end{definition}
The expected runtime is an expectation of number of steps after which we get the
result using Bernoulli process. Such approach has been used in
\cite{childs2004spatial,ambainis2005coins}, where the complexity was analysed.

Using the formalised description of quantum spatial search algorithms, we can
introduce the concept of \emph{attack on an algorithm} as follows.

\begin{definition}
	\emph{Attack on $\qss$} is a function $h$ such that 
	\begin{equation}
	h(\alg,t;\graph,\marked,\theta) = (\alg,t;\graph',\marked',\theta').
	\end{equation}
\end{definition}

It should be stressed that the attack cannot change the evolution model and
measure time of $\qss$. Still we will consider the function altering only
some of the elements of $\qss$. For example, the attacks restricted to graph structure
imply that $\marked=\marked'$ and $\theta=\theta'$. 

If we allow $\qss$ element to be changed, then we will say the element is
\emph{hackable}. Otherwise, it is not hackable. Note that by definition $\alg$
and $t$ are not hackable.

To quantify the efficiency of the attacks we introduce \emph{attack efficiency}
as follows.

\begin{definition}
\emph{Attack efficiency} $\eff_{h,\qss}$ on $\qss$ is defined as
\begin{equation}
\eff_{h,\qss} \coloneqq 1-\frac{\et_{\qss}}{\et_{h(\qss)}}.\label{eqn:attack-efficiency}
\end{equation}
\end{definition}
We are interested in such functions $h$ that $\eff_{h,\qss}\geq 0$. Furthermore,
since $t$ is common for both $\qss$ and $h(\qss)$ we have 
\begin{equation}
\eff_{h,\qss}=1-\frac{p(h(\qss))}{p(\qss)}.
\end{equation}

Let us consider the following scenario. The user is trying to start an algorithm
$\qss$, but the attacker has changed it into $\qss'$. If the user is aware of the attack and its nature, he can alter the
measurement time $t$ in order to minimize the expected runtime. To describe
this situation we introduce \emph{strong attack efficiency}.

\begin{definition} 
Suppose we have $\qss = (\alg,t;\graph,\marked,\theta)$ and $h:\qss \mapsto
(\alg,t;\graph',\marked',\theta')$. \emph{Strong attack efficiency}
$\overline{\eff}_{h,\qss}$ is defined as
\begin{equation}
\overline\eff_{h,\qss} \coloneqq
1-\frac{\et_{(\alg,t;\graph,\marked,\theta)}}{\min_{\tau\geq
		0}\et_{(\alg,\tau;\graph',\marked',\theta')}}. \label{eqn:strong-attack-efficiency}
\end{equation}
\end{definition}
This definition captures the best possible defence against
the attack.

All of the above definitions were restricted to the fixed spatial search
algorithm. Since our aim is to analyse the efficiency of attacks on more
general graph classes, we extend the previously defined terms.
\begin{definition}
Let $Q$ be a set of quantum spatial searches and $h$ be an attack on $Q$. Then
\emph{maximal attack efficiency on $Q$}, $\eff_{h,Q}$, is defined as
\begin{equation}
\eff_{h,Q} \coloneqq \max_{\qss \in Q}\eff_{h,\qss}.
\end{equation}
Similarly, \emph{maximal strong attack efficiency on $Q$}, $\eff_{h,Q}$, is defined as
\begin{equation}
\overline\eff_{h,Q} \coloneqq \max_{\qss \in Q}\overline \eff_{h,\qss}.
\end{equation}
\end{definition}
The above definition captures the pessimistic level of robustness against the attack.
Similarly, we can define the mean and minimal efficiencies. In practice, we
would like to find the dependence between the efficiency and the order of the
graph. 

Instead of providing deterministic $Q_n$, we will focus on $Q_n$ defined by
random graph models $\GG_n$. 
\begin{definition}
Let $\GG_n$ be a random graph model, and let for
arbitrary $\graph$ exist a quantum spatial search
$\qss(\graph)=(\alg,t(G);\graph,\marked(G),\theta(G))$. Let $h$ be an attack defined on $\qss$. We say that the
efficiency of the attack is almost surely at least $E_n$ iff
\begin{equation}
\PP({\eff}_{h,\qss(\graph)}\geq E_n| \graph\in \GG_n)
\xrightarrow{n\to\infty} 1.
\end{equation}
\end{definition}
The above definition can be naturally applied to strong efficiency. 

\section{Attacks on quantum spatial search algorithms}

\begin{figure}[t!]\centering
\subfigure[\label{fig:statistics-2}EC of order 2]{\includegraphics[]{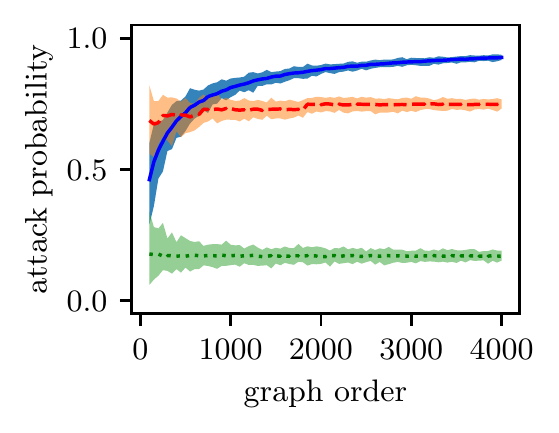}}
\subfigure[\label{fig:statistics-2or3}EC of order 2 or 3]{\includegraphics[]{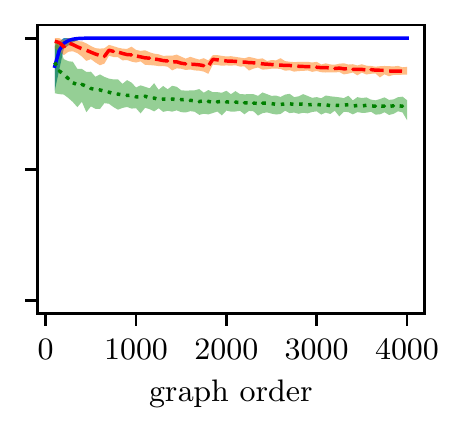}}
\subfigure[\label{fig:statistics-2or3close}EC of order 2 or 3 within distance 1]{\includegraphics[]{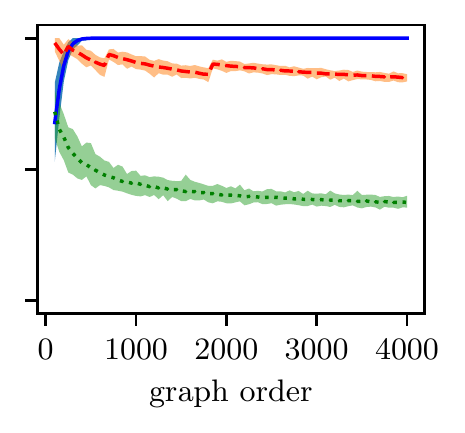}}
\caption{Probability of forming an exceptional configuration with a
        random vertex for different models of random graphs.
        \subref{fig:statistics-2} probability of forming EC of order 2;
        \subref{fig:statistics-2or3} probability of forming EC of order 2 or 3;
        \subref{fig:statistics-2or3close} probability of forming EC of order 2 or 3 within distance 1 from initially marked. For each graph order $n=100,150,\ldots,4000$ we have analyzed 100 graphs. The area describes the range of attack probability obtained from sampled graphs. Blue solid lines describe \ER graphs, orange dashed lines describe \WS graphs, and green dotted lines describe \BA graphs.}\label{fig:statistics}
\end{figure}

We consider the formation of exceptional configurations as a
method of attacking quantum spatial search.

We will consider the attacks altering the set of marked vertices only. For
quantum spatial search $(\textrm{SzQW},t;\graph,\{v\}, P_u)$, we add marked
vertices in such a way that the newly generated set  $S' \supset \{v\}$ forms a
connected exceptional configuration. As the probability of finding any vertex
from EC is much lower than the probability of finding a single marked vertex
\cite{nahimovs2017adjacent}, we decrease the success probability of the
algorithm.

Suppose that the algorithm outputs a vertex from $S'\setminus \{v\}$. Since we
choose additional marked vertices from close neighbourhood of the original 
marked vertex $v$, the possible defence against the attack is to make a simple
classical search over its neighbourhood, which is efficient for sparse graphs.
However, the cumulated success probability of measuring any of $S'$ is small.
Hence, the attack is effective.

Let us analyse \ER, \WS, and \BA models. For \ER we choose the probability of
adding edge $p=\frac{2\log(n)}{n}$, for \WS  we select initial degree $K=\lceil
2\log n\rceil $ and randomness $\beta =0.5$~\cite{watts1998collective}, and for
\BA we set attachment parameter to $m_0=3$~\cite{barabasi1999emergence,
	albert2002statistical}. First, we analyse the probability that for randomly
chosen vertex $v$ we can construct an exceptional configuration $S$ such that
$v\in S$. Next, we analyse the efficiency of the attack.

Numerical results demonstrate that for every vertex of \WS graph we can almost
surely find an exceptional configuration of order 3 (3EC), see
Fig.~\ref{fig:statistics}. Thus the model can be attacked almost surely for any
vertex. This is no longer the case when we allow constructing an exceptional
configuration of order 2 only.
 For \ER and \BA models, even 3EC will not provide
this kind of advantage for the attacker. Nevertheless, the probability of
attacking is still large, and the attacker may be able to construct an
exceptional configuration of a higher order.

Another constraint for resources available to the attacker is the distance between the originally marked vertices and the newly marked ones.
We consider two scenarios. For local exceptional configurations, the new vertices are direct neighbours of the original vertex. In the global scenario, they are at most neighbours of neighbours.
Based on Fig.~\ref{fig:statistics}, one can see that it is much easier to find global exceptional 
configurations than the local ones for \BA model. Hence the probability of 
attacking increases if the attacked has the ability to use such EC.
For \ER model the attacker cannot utilize 
this type of advantage, for given parametrization. Furthermore, for 
$p=\omega(\log(n)/n)$ all vertices have almost equal degrees 
\cite{bollobas1998random}. This makes finding path graph 3EC impossible, as 
degree condition from Preliminaries section cannot be fulfilled. We have not 
observed any dependence for \WS model.

%


\begin{figure}[t!]
	\centering\includegraphics[]{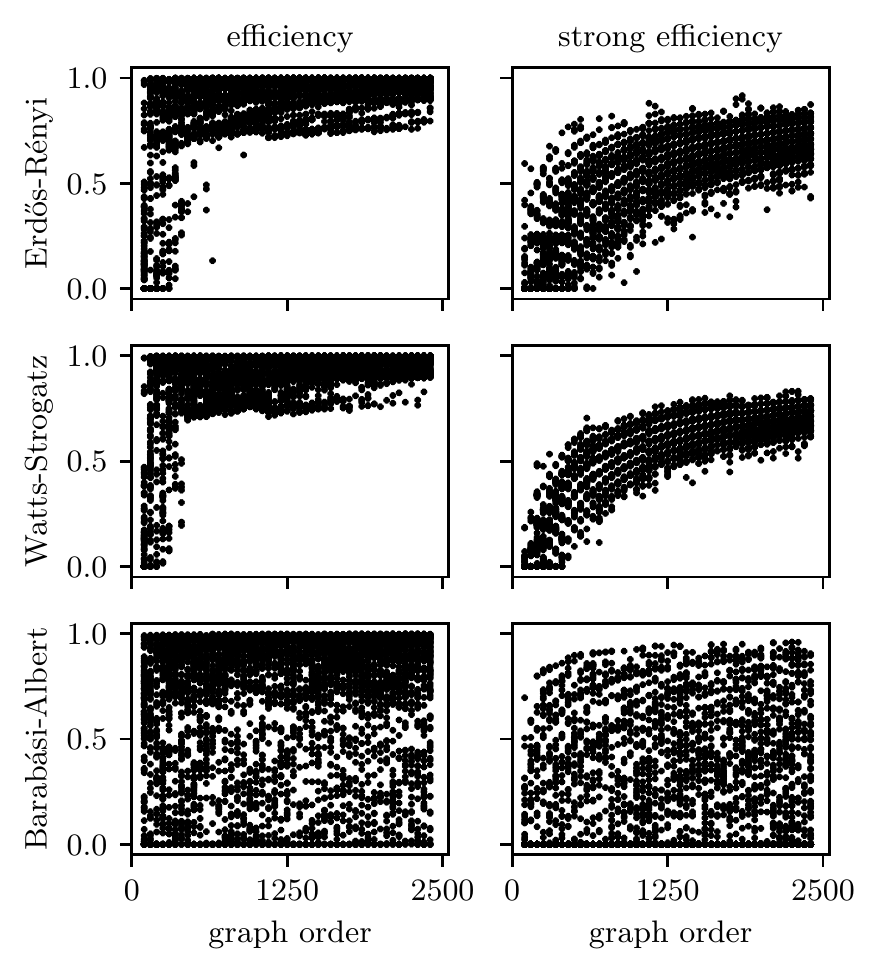}
	
	\caption{Attack efficiency and strong attack efficiency for random graph models. Pairs of vertices were chosen at random from the collection of all 2EC.
	Left plots present the efficiency (the case of unchanged measurement time). Right plots present the strong efficiency (the case with measurement time chosen
	optimally). While \ER and \WS present similar
	results, one can observe more robust behaviour for \BA
	model.}\label{fig:hacking-efficiency-ec}
\end{figure}

\begin{figure}[t!]
\centering\includegraphics{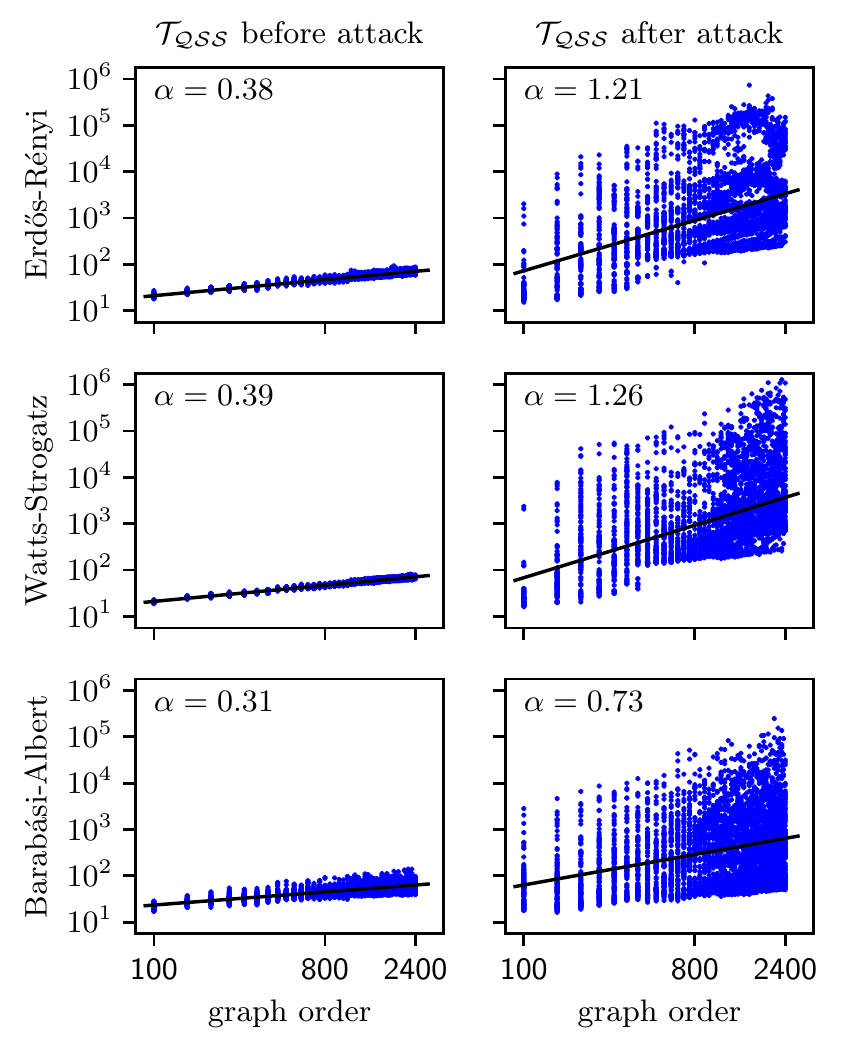}
\caption{Comparison of efficiency for the reference search (left plots) and the search after the attack (right plots) for different models of random graphs. Right plots provide detailed analysis of results concerning the efficiency case from Fig.~\ref{fig:hacking-efficiency-ec}.
The tangent of regression line, denoted by $\alpha$, is the numerically derived
complexity $\Theta(n^\alpha)$. One can observe  the increase in the run-time  resulting from the attack for \ER and \WS models.}\label{fig:ws-attack-ec-compare}
\end{figure}

For analysing the efficiency of the attack, we have chosen at random two 
vertices forming an exceptional configuration. We have analyzed 50 graphs for each order 
$n=100, 150,\dots, 2400$ for all models, using the optimization algorithm implemented in \pkgname{QuantumWalk} package~\footnote{Code available at \url{https://github.com/QuantumWalks/QuantumWalk.jl}} with 
penalty time $t_{\rm pen}=\lceil \log(n) \rceil$. Parameter $t_{\rm pen}$ is a 
value which is added to the time, thus changing the expected time into 
$(t+t_{\rm pen})/p$. Such adjustment prevents the optimization algorithm from 
halting at small time. Since $\log(n)$ is typically much smaller than $t$, its 
impact on our results is negligible.

In order to asses the impact of the attack on the complexity of quantum spatial search we have calculated the  expected runtime 
for three cases for each random graph. In the first case, which represents the reference search, we have a single marked element. This case provides the value of efficiency of the search $\mathcal{T}_{\qss}$ for a given graph. Next, we execute the $\qss$ on the graph modified by the addition of a single marked element, which results in the formation of an exceptional configuration. This case provides us the value of the increase of the expected run-time resulting from the attack, and thus the efficiency of the attack (cf.  Eq.~(\ref{eqn:attack-efficiency})). Finally, we calculate the expected run-time for the case when the user is aware of the attack and can utilize this knowledge to minimize its effects. In this case we calculate the strong efficiency (cf. Eq.~\ref{eqn:strong-attack-efficiency}).

Numerical results presented in Fig.~\ref{fig:hacking-efficiency-ec} show that
the attacking efficiency depends on the chosen graph model. Left three plots show that if user is not aware of the attack, the efficiency is almost surely 0.8 for \ER and almost surely 0.9 for \WS. For \BA the results are much more irregular, and no level of efficiency can be guaranteed. This suggests there are some other parameters influencing the efficiency of the attack.

If the user is aware of the attack, it is possible to prevent it at least
partially for all models by changing the measurement time. The efficiency in
this case is captured by the notion of \emph{strong  efficiency}. For \ER and
\WS models the strong efficiency is almost surely at least 0.5. For \BA we still
observe high irregularity in obtained data and the strong efficiency is
almost surely significantly smaller than 1.

As the complexity attack should result in the algorithm complexity growth, we
have determined numerically the expected time change in the case of common
measurement time. The results are presented in
Fig.~\ref{fig:ws-attack-ec-compare}. We have assumed that the complexity grows as the power complexity, $\Theta(n^\alpha)$. The tangent of the regression line of the expected run-time in the function of the graph order on the log-log scale provides the approximation of the parameter~$\alpha$. The observed growth of the values of $\alpha$, resulting from the attack, demonstrates that the attacker is able to significantly increase the expected run-time of the quantum algorithm. This suggests that the algorithm is vulnerable to the complexity attack, which might result in the denial-of-service of the quantum computer.

\section{Summary and discussion}

In this paper we have signified the problem of possible vulnerability of 
quantum algorithms to the complexity attacks. The presented approach is based 
on the analysis of input data. As such it can be used to discover 
weaknesses of quantum computers resulting from the application of quantum 
algorithms on input data unsuitable for processing on quantum 
machines. This is contrast to the common approach where only the theoretical 
computational complexity is taken into account.

We have developed the theoretical framework for quantifying the efficiency of 
the attacks. We have constructed an attack based on exceptional configurations 
and analysed it in the context of its applicability and efficiency. The 
analysis confirms that it is possible  to decrease the efficiency of quantum 
spatial search based on Szegedy walk by malicious modification of input data. 

One should note that the presented results can be applied for a general 
class of graphs. This is in contrast to the results from 
\cite{nahimovs2017adjacent}, where only special classes of graphs were 
considered. For those classes it can be shown analytically that the algorithm 
complexity changes from $\Theta(\sqrt{n})$ to $\Theta(n)$.

It should be stressed that the models of random graphs used for assessing the
security of quantum algorithms mimic the structure of real-world
data~\cite{albert2002statistical}. As such the presented analysis confirms that
the theoretical security of quantum procedures can be inadequate when the
algorithms are applied for specific input data. This includes input data which
encode the connections observed in complex networks.

We should also stress that the scenario considered in this work does not focus
on the security of the communication which takes place between the client and
the server. On the contrary, our figure of merit is the impact of the modified
data on the  computational resource used by the server. This is in contrast to
the protocols based on the concept of Universal Blind Quantum
Computation~\cite{broadbent2009universal}. Such protocols enable the detection
of the input data modification, if the data are encoded in a quantum state.
Alice can use such protocols to counteract the effect of data alteration by
informing Bob about the Eve's activity. However, such protocol cannot be used to
prevent Alice from sending malformed input data. On the other hand in our
scenario it is easy to include the possibility that Alice is responsible for
sending data which are already malformed -- either on purpose or by accident --
and our goal is to asses how often such situation can occur when we consider
data modeled by complex networks. Our results show that secure transmission is
insufficient for ensuring the availability of the quantum hardware.

\paragraph{Acknowledgments} The authors would like to thank Nikolay Nahimov for
discussion concerning exceptional configurations and Alexander Rivosh for
inspiring the application of this concept in quantum spatial search, and Iza
Miszczak for reviewing the manuscript. This work has been partially supported by
Polish National Science Center grant 2011/03/D/ST6/00413. Numerical calculations
were possible thanks to the support of PL-Grid Infrastructure. We would also
like to thank the anonymous reviewer for his comments, which greatly improved
the quality of our paper.

\bibliographystyle{ieeetr}
\bibliography{quantum_spatial_search_attack}
	
\end{document}